\documentclass[11pt,a4paper,reqno]{amsart}
\usepackage{amsthm,amsmath,amsfonts,amssymb,amsxtra,appendix,bookmark,dsfont,latexsym,epsfig,graphicx,stix}

\makeatletter
\usepackage{hyperref}
\usepackage{delarray,color}
\usepackage{euscript}

\usepackage{hyperref}

\makeatletter
\def\@tocline#1#2#3#4#5#6#7{\relax
  \ifnum #1>\c@tocdepth % then omit
  \else
    \par \addpenalty\@secpenalty\addvspace{#2}%
    \begingroup \hyphenpenalty\@M
    \@ifempty{#4}{%
      \@tempdima\csname r@tocindent\number#1\endcsname\relax
    }{%
      \@tempdima#4\relax
    }%
    \parindent\z@ \leftskip#3\relax \advance\leftskip\@tempdima\relax
    \rightskip\@pnumwidth plus4em \parfillskip-\@pnumwidth
    #5\leavevmode\hskip-\@tempdima
      \ifcase #1
       \or\or \hskip 1em \or \hskip 2em \else \hskip 3em \fi%
      #6\nobreak\relax
      \dotfill
      \hbox to\@pnumwidth{\@tocpagenum{#7}}
    \par
    \nobreak
    \endgroup
  \fi}
\makeatother

\newcommand{\bdm}{\begin{displaymath}}
\newcommand{\edm}{\end{displaymath}}
\newcommand{\bdn}{\begin{eqnarray}}
\newcommand{\edn}{\end{eqnarray}}
\newcommand{\bay}{\begin{array}{c}}
\newcommand{\eay}{\end{array}}
\newcommand{\ben}{\begin{enumerate}}
\newcommand{\een}{\end{enumerate}}
\newcommand{\R}{\mathbb{R}}
\newcommand{\C}{\mathbb{C}}

\newcommand{\cL}{\mathcal{L}}
\newcommand{\PsiLau}{\Psi_{\rm Lau}}
\newcommand{\PsiLaun}{\Psi_{\rm Lau} ^{(\ell)}}
\newcommand{\cLau}{c _{\rm Lau}}

\newtheorem{theorem}{Theorem}[section]

\newtheorem{conjecture}[theorem]{Conjecture}

\theoremstyle{remark}

\newcommand{\beq}{\begin{equation}}
\newcommand{\eeq}{\end{equation}}
\numberwithin{equation}{section}

\newcommand{\bx}{\mathbf{x}}
\newcommand{\by}{\mathbf{y}}
\newcommand{\im}{\mathrm{i}}
\newcommand{\1}{\mathds{1}}

\newcommand{\LLLf}{\mathrm{LLL}_{\rm asym} ^N}

\def\XXint#1#2#3{{\setbox0=\hbox{$#1{#2#3}{\int}$}
     \vcenter{\hbox{$#2#3$}}\kern-.5\wd0}}

\begin{document}

\title{On the stability of Laughlin's fractional quantum Hall phase}

\author[N. Rougerie]{Nicolas Rougerie}
\address{Ecole Normale Sup\'erieure de Lyon \& CNRS,  UMPA (UMR 5669)}
\email{nicolas.rougerie@ens-lyon.fr}

% \affiliation{Universit\'e Grenoble-Alpes \& CNRS ,  LPMMC (UMR 5493), B.P. 166, F-38 042 Grenoble, France}
% %\email{nicolas.rougerie@lpmmc.cnrs.fr}

% \author{Jakob Yngvason}
% \address{Faculty of Physics, University of Vienna, Boltzmanngasse 5, A-1090 Vienna, Austria}
% \email{jakob.yngvason@univie.ac.at}

% \affiliation{Faculty of Physics,
% University of Vienna, Boltzmanngasse 5, A-1090 Vienna, Austria}
%\email{jakob.yngvason@univie.ac.at}

%\affiliation{Erwin Schr{\"o}dinger Institute for Mathematical Physics, Boltzmanngasse 9, A-1090 Vienna, Austria}

\date{August, 2022}

\begin{abstract}
The fractional quantum Hall effect in 2D electron gases submitted to large magnetic fields remains one of the most striking phenomena in condensed matter physics. Historically, the first observed signature is a Hall resistance quantized to the value $h/(e^2\nu)$ when the filling factor $\nu$ (electron density divided by magnetic flux quantum density) of a 2D electron gas is in the vicinity of an inverse odd integer $\nu \approx 1/(2m+1)$. This was one of the first observation of fractional quantum numbers.  A large part of our basic theoretical understanding of this effect (and descendants) originates from Laughlin's theory of 1983, reviewed here from a mathematical physics perspective. We explain in which sense Laughlin's proposed ground and excited states for the system are rigid/incompressible liquids, and why this is crucial for the explanation of the effect.   
\end{abstract}

\maketitle

\begin{center}
 \emph{This essay is intended as a contribution to the second edition of the \emph{Encyclopedia of condensed matter physics}. It is partially based on two previous review texts:~\cite{Rougerie-xedp19,Rougerie-Elliott}.}
\end{center}

\tableofcontents

\newpage

\section{Key objectives}

\begin{itemize}
 \item Discuss the basic phenomenogy of the fractional quantum Hall effect (FQHE) in 2D electron gases (2DEG). 
 \item Introduce Laughlin's theory of the effect at filling fraction $1/\ell$, $\ell$ an odd integer.
 \item Highlight two key incompressibility/rigidity properties the theory relies on.
 %\item Report on recent mathematically rigorous progress on said properties. 
 \item Explain the rigorous derivation of Haldane pseudo-potentials (whose ground eigenstates are generated from Laughlin's function) from first principles. 
 \item State the (still open) spectral gap conjecture for Haldane pseudo-potentials (first key rigidity property).
 \item State incompressibility estimates ensuring that the Laughlin phase is stable against external potentials and residual interactions (second key rigidity property).
\end{itemize}

\section{Phenomenogy of the fractional quantum Hall effect}

\subsection{Experimental facts}

The (fractional) (quantum) Hall effect~\cite{Jain-07,Girvin-04,Goerbig-09,StoTsuGos-99,Laughlin-99} concerns the charge transport properties in 2D samples submitted to large magnetic fields. The Lorentz force exerted by the latter on moving charges leads to a non-trival transverse resistance $R_{xy}= V_y/I_x$ when a current $I_x$ is applied in some direction $x$. The classical effect has historically served as a way of measuring the charge carriers' density in a given sample. It is however in the 1980's that dramatic, purely quantum signatures were discovered in this context: the quantum Hall effect (integer, then fractional). This lead to two Nobel prizes in physics (von Klitzing 1985, St\"ormer-Tsui-Laughlin 1998) and a half (Thouless-Haldane-Kosterlitz 2016) and the advent of topology as a tool to classify phases in condensed matter physics.

 \begin{figure}\label{fig:FQHE}
\includegraphics[width=10cm]{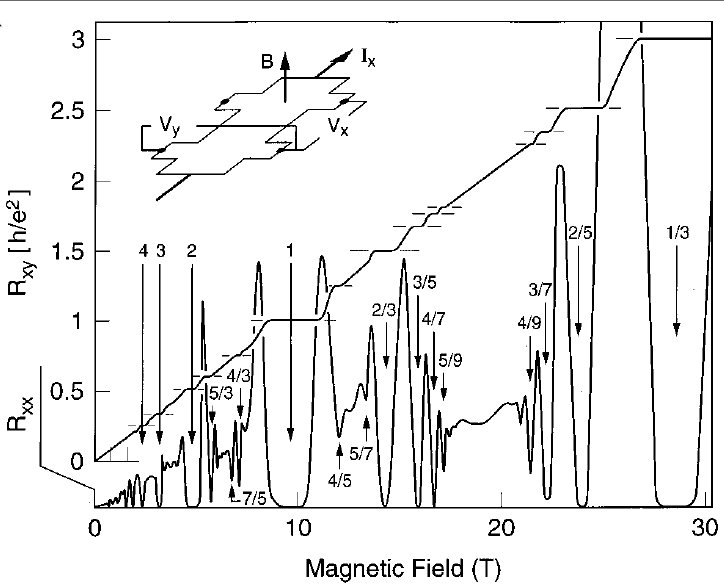}
\caption{The fractional quantum Hall effect~\cite{StoTsuGos-99}. Sketch of the experimental sample in top-left corner. Plots of the longitudinal $R_{xx} = V_x /I_x$ and transverse (Hall) $R_{xy} = V_y/I_x$ resistances as a function of the magnetic field.}
\end{figure}

We will limit our discussion to (some of the) most striking experimental findings as depicted on Figure~$1$. Namely, consider a 2D gas of electrons submitted to a current and let the filling factor 
\begin{equation}\label{eq:filling}
\nu := \frac{hc}{e} \frac{\rho}{B} 
\end{equation}
with $\rho$ the electrons' density, $B$ the applied magnetic field and $h,c,e$ respectively Planck's constant, the speed of light and the elementary charge. Around certain particular values of said factor:

\begin{itemize}
 \item The direct resistance $R_{xx} = V_x /I_x$ exhibits sudden drops to almost $0$ values.
 \item Simultaneously the Hall/transverse resistance $R_{xy} = V_y/I_x$ is very precisely quantized to the value $h/(e^2\nu)$, which stays stable for a certain window (plateau) of the applied field/filling factor. 
\end{itemize}
Note that the value $h/(e^2\nu)$ taken by $R_{xy}$ is precisely that one can derive from classical considerations. Thus the essence of the quantum effect is the plateau: that the measured resistance sticks on this value for a finite window of $\nu$'s around certain particular values.

The special values of $\nu$ at which the above happens are 

\begin{itemize}
 \item integers $\nu = 1,2,3 \ldots$. This is the integer quantum Hall effect (IQHE). 
 \item certain fractions, in particular $\nu = 1/m, m$ odd. This is the fractional quantum Hall effect (FQHE). 
\end{itemize}
The classification of all fractions at which the effect should occur is not a closed topic (as far as I know) but the most prominently observed are of the form 
 \begin{equation}\label{eq:Jain frac}
 \nu = m + \frac{p}{2pn + 1}, \quad m,p,n  \mbox{ integers}. 
\end{equation}
This is certainly the case on Figure~$1$, where actually one mostly sees the case $n=1$.  The particular case $p=1,n$ integer corresponds to Laughlin fractions, discussed below. For larger $p$ one gets Jain fractions ($n=1,p$ integer corresponds to the principal Jain sequence, most prominent on the figure), explained in terms of the composite fermions theory~\cite{Jain-07}, a generalization of Laughlin's picture we will not touch upon. Fractions of the form 
$$ m + 1 - \frac{p}{2pn + 1}$$
correspond to a certain particle-hole transformation of those of the form~\eqref{eq:Jain frac}, and their theory is thus the same. 

Laughlin's theory and its composite fermions generalization give a rationale for essentially all the features observed on Figure~$1$. The most noteworthy unclear feature lies in the oscillations\footnote{Something different occurs at $\nu = 5/2$, see e.g.~\cite{Yngvason-22}.} in $R_{xx}$ around $\nu = m + \frac{1}{2}, m= 0,1$. Those however also have an explanation in terms of the composite fermions theory~\cite{Jain-07}, that we will shall not discuss.

\subsection{Theoretical road-map}

We will in the sequel give a mathematical physics perspective on the above facts, taking for granted the generally accepted hierarchization of energy scales leading to the effect, at least in its purest form:
\begin{enumerate}
 \item The perpendicular magnetic field is so large that the magnetic kinetic energy of electrons is by and large the main player.
 \item Next comes the repulsive interaction energy of electrons, due to Coulomb forces. The short-range, singular, part is thought to be the most important. 
 \item Finally all other energies are small compared to the previous ones. In particular, the temperature is neglected altogether. However, the electrostatic potential generated by impurities in the sample is crucial to the effect, and must be taken into account.
\end{enumerate}

The essence of Laughlin's theory is that it provides a tentative ground state/vacuum for the system so that 
\begin{itemize}
 \item The magnetic kinetic energy is minimized exactly. 
 \item The interaction energy is strongly reduced, in particular its short range part. 
 \item The filling factor is close to $1/m$, $m$ an odd integer. This appears a posteriori after considering the first two points. 
 \item The shape of the ground state is very robust, in particular in its response to residual interactions and/or external fields. 
 \item The response to external fields is to generate quasi-particles/holes of charge $e/m$, which serve as effective charge carriers in transport experiments.
\end{itemize}

The fourth point in particular is the aspect refered to in the title of this essay.

\section{Basic theory}\label{sec:FQHE}

Before going into more precise statements regarding the rigidity/incompressibility of the Laughlin state, we explain its basic, heuristic, derivation.

\medskip

\noindent\textbf{The many-body quantum Hamiltonian}. We start from a basic Hamiltonian for the quantum 2D electron gas (in adimentionalised form $\hbar = c = e = 2m = 1$)
\begin{equation}\label{eq:start hamil}
H_N^{\rm QM} = \sum_{j=1} ^N \left[\left( -\im \nabla_{\bx_j} - \frac{B}{2} \bx_j ^\perp \right) ^2 + V (\bx_j) \right]+ \sum_{1\leq i < j \leq N} W(\bx_i-\bx_j) 
\end{equation}
acting on $L_{\rm asym} ^2 (\R ^{2N}) $, the Hilbert space for $N$ 2D fermionic particles. Here $\bx^{\perp}$ denotes the vector $\bx\in \R^2$ rotated by $\pi/2$ counter-clockwise, so that 
$$ \mathrm{curl} \frac{B}{2} \bx ^\perp = B$$
and thus $\frac{B}{2} \bx ^\perp$ is the vector potential of a uniform magnetic field, expressed in symmetric gauge. In view of our choice of units, $B$ is actually $\sqrt{\alpha}$ times the physical magnetic field, with $\alpha = e^2 /(\hbar c) \sim 1/137$ the fine structure constant, see e.g.~\cite[Section~2.17]{LieSei-09}. 

We take into account an external potential $V:\R^2 \mapsto \R$ modeling trapping and/or impurities in the sample, and repulsive pair interactions $W:\R^2 \mapsto \R$ between particles. Typically $W$ should be the 3D Coulomb kernel (with $\alpha$ the fine structure constant again)
\begin{equation}\label{eq:Coul}
 W (\bx-\by) = \frac{\alpha}{|\bx-\by|} 
\end{equation}
or some screened version. We have made the customary assumption that the magnetic field is strong enough to polarize all the electrons' spins.

\bigskip

\noindent\textbf{Quantum Hall plateaux.}~The extremely precise quantization to particular values of $R_{xy}$ (read on the vertical axis of Figure~$1$) has an interpretation in terms of topological invariants of the system~\cite{BelSchEls-94,Frohlich-92,Frohlich-95}, but that is not what we focus on here. Instead, looking at the horizontal axis of Figure~$1$, we see that the particular features occur around special values (the numbers associated with arrows on the picture) of the filling factor~\eqref{eq:filling} of the system. 

In the sequel we (partially) address only the question ``\emph{why} does something special happen at these parameter values~?'' without touching much on the ``\emph{how} does the particular observed experimental signature emerge~?'' In a nutshell, the integer values found for $R_{xy}^{-1}$ in the IQHE are Chern numbers associated to the ground state of free electrons in large magnetic fields. The fractional values of the FQHE can roughly be thought of as Chern numbers associated to the ground states of free quasi-particles generated on top of the strongly correlated FQH ground states.

\bigskip

\noindent\textbf{Landau levels.} The workhorse of the quantum Hall effect is the quantization of kinetic energy levels in the presence of a magnetic field. Namely, the appropriate kinetic energy operator for a 2D particle in a perpendicular magnetic field $B$ is 
\begin{equation}\label{eq:Landau}
 H = \left( -\im \nabla_{\bx} - \frac{B}{2} \bx ^\perp \right) ^2 
\end{equation}
acting on $L^2 (\R^2)$. 

The energy levels (eigenvalues) of the above are well-known~\cite{RouYng-19,Jain-07} to be $2B (n+1/2)$ for integer $n$, since one can write 
$$ H = 2B \left( a^\dagger a + \frac{1}{2}\right)$$
for appropriate ladder operators $a,a^\dagger$ with $[a,a^\dagger] = 1$. The lowest eigenspace (lowest Landau level, corresponding to the eigenvalue $B$) can be represented as 
\begin{equation}\label{eq:LLL}
\mathrm{LLL} = \left\{ f(z) e^{-\frac{B}{4}|z|^2} \in L ^2 (\R^2), f \mbox{ holomorphic }\right\}
\end{equation}
and the $n$-th Landau level can be obtained as $\left(a^\dagger\right)^n \mathrm{LLL}.$ Hence each energy level is infinitely degenerate when working on the full plane. Well-known arguments indicate that this degeneracy is reduced in finite regions, with a degeneracy $\propto B \times \mbox{ Area }$. One argument for this is that~\eqref{eq:Landau} can be restricted to a rectangle whose area is a multiple of $2\pi B^{-1}$, imposing magnetic-periodic boundary conditions see~\cite{AftSer-07,Almog-06,FouKac-11,Perice-22,NguRou-22} or~\cite[Sections~3.9 and 3.13]{Jain-07}. The energy levels are then the same as above, with degeneracy exactly $B (2\pi)^{-1} \times$ area of the rectangle.

\bigskip

\noindent\textbf{The integer quantum Hall effect.}~Some plateaux (left of Figure $1$) in $R_{xy}$/drops in $R_{xx}$ occur at integer values of $\nu$ and it is not surprising that something special should happen there (again, it is highly non-trivial to derive the specific signature of the ``something special''). This can be understood in a non-interacting electrons picture, taking only the Pauli exclusion principle into account. One assumes that the magnetic kinetic energy, proportional to $B$, is the main player and that all other energy scales in~\eqref{eq:start hamil} are negligible against it. By this we mean that $W$ is dropped in~\eqref{eq:start hamil} and that the only effect of $V$ is to essentially confine the gas to a domain $\Omega$.

As the name indicates, the filling factor measures the ratio of electron number to number of available one-body states $N_B (\Omega)$ in a given Landau level (see the above considerations, keeping in mind that $(2\pi)^{-1} h = c = e = 1$): 
$$ \nu = 2\pi \frac{\rho}{B} = 2\pi \frac{N}{|\Omega| B} \simeq \frac{N}{N_B (\Omega)}$$
if $N$ electrons are confined to the region $\Omega$ with density $\rho = N/|\Omega|$. In the ground state of an independent electron picture (taking only the Pauli exclusion principle into account), one fills the eigenstates of~\eqref{eq:Landau} with one electron each, starting from the lowest one. At integer $\nu$, the $\nu$ lowest Landau levels are thus completely filled, and the others completely empty, a very rigid and non-degenerate situation. This rigidity is actually important in order to treat the energy scales other than $B$ perturbatively.  

\bigskip

\noindent\textbf{The fractional quantum Hall effect.}~Many plateaux however occur at particular \emph{rational} filling factors and are impossible to explain in an independent electrons picture. Laughlin's groundbreaking theory~\cite{Laughlin-83,Laughlin-87,Laughlin-99} explains why something special ought to occur at 
\begin{equation}\label{eq:Lau frac}
 \nu = \frac{1}{\ell}, \quad \ell \mbox{ an odd integer} 
\end{equation}
e.g. at the right-most plateau $\nu = 1/3$ of Figure~$1$, but also at $\nu = 1/5$, a fraction also observed in experiments ($\nu = 1/9$ and lower is not observed, while $\nu = 1/7$ is borderline). The $\nu = 1/3$ fraction is the first to have been observed~\cite{StoTsuGos-82}, and the most stable. There are other, more exotic, fractions and features, but let us not get into that to focus on Laughlin's theory of the mother of all fractions, namely~\eqref{eq:Lau frac}.

\bigskip

\noindent\textbf{Restriction to the lowest Landau level}. We henceforth restrict to filling factors $\nu < 1$ . In the regime relevant to the quantum Hall effect, the gap $B$ between the magnetic kinetic energy levels is so large that the first approximation we make is to project all the physics down to as few Landau levels as possible. With filling ratio $\nu \leq 1$, the lowest Landau level is vast enough (again, see the above heuristics) to accommodate all particles, and thus we restrict available many-body wave-functions to those made entirely of lowest Landau\footnote{Generalizations to larger filling factors, when one works in an excited Landau level, are discussed in~\cite{RouYng-19}.} levels orbitals~\eqref{eq:LLL}. It is in fact convenient to work on the full space at first. The restrictions to finite area/density will actually be performed later, and we will have to make sure they are coherent with our aim: a thermodynamically large system with density $\rho \sim B \nu (2\pi)^{-1}$.

\bigskip

\noindent\textbf{Killing the interaction's singularity}. The main energy scale, the magnetic kinetic energy, is now frozen by projecting all one-body states to~\eqref{eq:LLL}. Laughlin's key idea is that the next energy scale to be considered is the pair interaction, and more precisely its singular short-range part. Any tentative ground state ought to belong to
\begin{equation}\label{eq:LLLN}
 \mathrm{LLL}_N = \left\{ A(z_1,\ldots,z_N) e^{-\frac{B}{4} \sum_{j=1} ^N |z_j| ^2 }, \quad A \mbox{ analytic and antisymmetric} \right\}
\end{equation}
and, for $\ell$ odd,  the wave-function
\begin{equation}\label{eq:PsiLau}
 \PsiLau ^{(\ell)} (z_1,\ldots,z_N):= \cLau \prod_{1\leq i < j \leq N} (z_i-z_j) ^\ell e^{-\frac{B}{4}\sum_{j=1} ^N |z_j| ^2}
\end{equation}
is introduced in order to reduce as much as possible the probability of particle encounters. $\PsiLau^{(\ell)}$ is designed to vanish when $z_i=z_j$ while preserving the anti-symmetry and analyticity. It may seem that $\ell$ is a free variational parameter. But so far we thought somewhat grand-canonically: we have not fixed the density of our system yet. It turns out that the one-particle density of Laughlin's function satisfies
\begin{equation}\label{eq:dens Lau}
 \varrho_{\PsiLau} (\bx) \simeq \frac{B}{2\pi \ell} \1_{|\bx| \leq \sqrt{\frac{2N\ell}{B}}}.
\end{equation}
That is, it lives on a thermodynamically large length scale (whose disk shape shall not bother us to determine bulk properties)  and has filling factor $\nu = \ell^{-1}$ (recall the choice of units in~\eqref{eq:start hamil}). This can be proved rigorously, see e.g.~\cite{RouSerYng-13a,RouSerYng-13b} and references therein. A common hand-waving heuristic is that in the construction of $\PsiLau ^{(\ell)}$, one needs $\ell N$ single particle orbitals 
$$\varphi_k (z)= z^k e^{-\frac{B}{4} |z| ^2}.$$
Hence the ratio of particle number to avaible states discussed above should indeed be $\ell ^{-1}.$ One can also derive~\cite{BerHal-08} that the occupation number of each orbital is $\sim \ell ^{-1}$.

Now we can answer our original question ``what is special about filling factor $\nu = \ell ^{-1}$ ?'' The answer is that, at such parameter values, we may form a Laughlin state of exponent $\ell$ as approximate ground state of our system. It minimizes the magnetic kinetic energy exactly, and does a very good job at reducing the short-range part of the interaction.
% 
% \bigskip
% 
% \noindent\textbf{Towards a rigorous derivation of Laughlin's function.} The second part of the above derivation is very heuristic, and will probably stay that way in the case of the true 3D Coulomb interaction. However, if one is willing to approximate the short-range part of the interaction as a sharply peaked delta-like potential, one may indeed derive rigorously the Laughlin state (and/or variants) in a physically relevant limit. This is based on the fact that the Laughlin function is an \emph{exact} ground state for an approximate interaction of zero-range, projected on the lowest Landau level. For a precise formulation of this, and the derivation of such model interactions from scaled ones, I refer to~\cite{LewSei-09,SeiYng-20}. A Gross-Pitaevskii-like limit of bosonic models based on effective delta interactions projected in the lowest Landau level is studied in a very nice paper by Elliott and co-workers~\cite{LieSeiYng-09}.
% 
% The main open problem in this direction is to make the derivation of Laughlin's function alluded to above \emph{uniform in the particle number} $N$. This depends on a \emph{spectral gap conjecture} for effective zero-range interactions, whose formulation can be found in~\cite[Appendix]{Rougerie-xedp19} and references therein. Partial progress towards the conjecture are in~\cite{NacWarYou-20a,NacWarYou-20b,WarYou-21,WarYou-21b}.

\bigskip

\noindent\textbf{Laughlin quasi-holes}. So far we have argued that Laughlin's function is a good ansatz for the ground state of the system at the relevant filling factor, when neglecting the effect of the external potential $V$ and the long-range part of the interaction $W$ in~\eqref{eq:start hamil}. That is not the end of the story, for the latter ingredients do exist in actual experiments, in particular, the disorder landscape that impurities enforce in $V$ is crucial to the quantum Hall effect. It leads to the finite width of the plateau by localizing charge carriers generated when the filling factor varies in the vicinity of a stable (incompressible) fraction~\cite{Laughlin-81}.

The Laughlin state should in fact be seen as the ``vacuum'' of a theory explaining the FQHE experimental data. The next step is to construct the quasi-particles generated from said vacuum when suitably moderate external fields are applied, such as those generating the currents in experiments. 

It is in fact easier to argue about quasi-holes, generated e.g. when the filling factor is lowered a little from the magic fraction $\ell ^{-1}$, as when moving towards the right on Figure~$1$. The salient feature is that we stay on the same FQHE plateau for a while when doing so. It must hence be that the ground state of the system stays ``Laughlin-like'' for reasonably smaller $\nu$. In fact, Laughlin's next key idea is two-fold: 
\begin{itemize}
 \item for smaller filling factors, the ground state is generated from~\eqref{eq:PsiLau} by adding uncorrelated quasi-holes. These are typically pinned by the impurities of the sample (modeled by $V$ in~\eqref{eq:start hamil}).
 \item when applying an external field at $\nu$ close to $\ell^{-1}$, the current is carried by the motion of such quasi-holes. 
\end{itemize}
The second idea in particular is quite far-reaching: it has by now been measured~\cite{SamGlaJinEti-97,YacobiEtal-04,MahaluEtal-97} that the current is carried in fractional lumps of $e \ell^{-1}$ and~\cite{BarEtalFev-20,NakEtalMan-20} that the charge carriers obey fractional quantum statistics, i.e. are emergent anyons~\cite{AroSchWil-84,Halperin-84,LunRou-16,LamLunRou-22,ZhaSreGemJai-14,CooSim-15}.

To give a bit more mathematical flesh to these heuristics, observe that our considerations above (minimization of the magnetic kinetic energy, almost minimization of the interaction energy) generally suggest to look for states of the form 
\begin{equation}\label{eq:PsiF}
 \Psi_F (z_1,\ldots,z_N):= c_F \PsiLaun (z_1,\ldots,z_N) F (z_1,\ldots,z_N)
\end{equation}
with $F$ analytic and symmetric, $c_F$ a $L^2$-normalization constant. The next key steps has a ``why go for complications if we can try something simple first'' flavor. Namely we consider only a subset of the above possible states, those of the form
\begin{equation}\label{eq:Psif}
 \Psi_f (z_1,\ldots,z_N):= c_f \PsiLaun(z_1,\ldots,z_N) \prod_{j=1} ^N f(z_j)
\end{equation}
where $f:\C \mapsto \C$ is analytic and $c_f$ is a normalization constant. In some sense, we try not to add extra correlations on top of the already strongly correlated $\PsiLaun$.

It turns out that states of the form~\eqref{eq:Psif} give sufficient freedom to explain the effect. Namely, since $f$ is essentially a polynomial, we write it in the manner 
\begin{equation}\label{eq:poly}
f(z) = \prod_{k=1} ^K (z-a_k) 
\end{equation}
for points $a_1,\ldots,a_K \in \C$. Since $\Psi_f$ must vanish whenever any of the electrons coordinates approaches some $a_k$, those are interpreted as the (here, classical) locations of quasi-holes, whose role in the effect we discussed above. 

\medskip

\noindent\textbf{Stability of the Laughlin phase}. In the next section we discuss what is known/hoped for at a mathematical physics level of precision regarding two assumptions implicitly made above:
\begin{itemize}
 \item The space of functions of the form~\eqref{eq:PsiF} is indeed an approximate ground eigenspace for (at least the singular part of the) the interaction energy. It is separated from the rest of the spectrum by an energy gap, so that remaining energy scales can be treated perturbatively. 
 \item When minimizing the remaining energy scales in the space~\eqref{eq:PsiF} (in the spirit of degenerate perturbation theory), it is legitimate to restrict to the simpler form~\eqref{eq:Psif} of Laughlin plus quasi-holes wave-functions.
\end{itemize}
These two aspects are manifestations of the Laughlin state's rigidity/incompressibility. In fact the first one is most often refered to as incompressibility, so that we will refer to the second one as rigidity.
 
\section{Mathematical results and conjectures} 

We now discuss in more mathematical details the two questions we mentioned last: (i) that the space~\eqref{eq:PsiF} constructed from Laughlin's function almost minimizes the interaction energy, (ii) that the subset of Laughlin-plus-quasi-holes functions~\eqref{eq:Psif} is a stable subset of~\eqref{eq:PsiF}.

\subsection{Haldane pseudo-potentials}\label{sec:gap}

In the case of true interactions, e.g. Coulombic~\eqref{eq:Coul}, the Laughlin function is  a good guess, but there is no obvious way of justifying this in a well-defined/controled limit/approximation. However, the question can be given a clear mathematical meaning modulo simplifying the true interaction.

Namely, consider a toy Hamiltonian defined as follows. Let the fermionic lowest Landau level be 
\begin{align}
%\LLLb &= \left\{ A(z_1,\ldots,z_N) e^{-\frac{B}{4} \sum_{j=1} ^N |z_j| ^2 }, \quad A \mbox{ analytic and symmetric} \right\} \\
\LLLf &= \left\{ A(z_1,\ldots,z_N) e^{-\frac{B}{4} \sum_{j=1} ^N |z_j| ^2 }, \quad A \mbox{ analytic and antisymmetric} \right\}
\end{align}
where antisymmetric means ``under exchange of the labels of the coordinates $z_1,\ldots,z_N$''. On this space, consider the action of the $m$-th Haldane pseudo-potential Hamiltonian 
\begin{equation}\label{eq:pseupot}
H (m,N) := \sum_{1\leq i < j \leq N} |\varphi_m \rangle \langle \varphi_m |_{ij} 
\end{equation}
where $|\varphi_m \rangle \langle \varphi_m |_{ij}$ projects the relative coordinate\footnote{We identify points in the plane with complex numbers} $\bx_i - \bx_j$ of particles $i$ and $j$ on the one-body state ($c_m$ is a normalization constant)
$$\varphi_m (z) = c_m z^m e^{-\frac{B}{4} |z| ^2}.$$
Note that, when acting on $\LLLf$, only for odd $m$ does $H(m,N)$ act non-trivially.

To motivate the above definition, we recall that the magnetic kinetic energy is the main energy scale, with discrete energy levels separated by huge gaps. Perturbation theory tells us that we should look for the ground state of the system by minimizing, in the ground eigenspace $\LLLf$, the next main energy scale, namely the interaction. If one projects a bona-fide pair interaction Hamiltonian
$$ H_w = \sum_{1\leq i< j \leq N} w (x_i-x_j)$$
with \emph{radial} potential $w\geq 0$ on the LLL, one obtains 
\begin{equation}\label{eq:LLL hamil}
H_w ^{\mathrm{LLL}} := P_{\mathrm{LLL}^N_{\rm sym/asym}} H_w \: \: P_{\mathrm{LLL}^N_{\rm sym/asym}}= \sum_{i<j} \sum_{m\geq 0} \left\langle \varphi_m | w | \varphi_m \right\rangle |\varphi_m \rangle \langle \varphi_m |_{ij}. 
\end{equation}
The coefficients $\left\langle \varphi_m | w | \varphi_m \right\rangle$ are called ``Haldane pseudo-potentials''~\cite{Haldane-83,BerPap-99,RouSerYng-13b,LieSeiYng-09,LewSei-09,PapBer-01,GirJac-84,Viefers-08}. The toy Hamiltonian~\eqref{eq:pseupot} above is obtained by discarding all terms from the sum but one, in order for the Laughlin state to be an exact ground state, and not just a very good approximation. Indeed, $\PsiLaun$ is clearly an exact ground state
$$ H(\ell - 2,N) \PsiLau^{(\ell)} = 0.$$

The rigorous justification of the expansion in Haldane pseudo-potentials and the truncation of the series is considered in~\cite{LewSei-09,SeiYng-20} (with techniques whose inspiration goes to back to~\cite{Dyson-57}, see~\cite{LieSeiSolYng-05,Rougerie-EMS}) in the limit of strong short-ranged potentials. One must be careful that for such singular potentials, the Haldane pseudo-potentials have to be modified to account for short-range correlations due to usual two-body scattering. This involves states outside the lowest Landau level.

Let $a>0$ be a (small) length and (note that we subtract the $\mathrm{LLL}$ ground state energy for convenience)
\begin{equation}\label{eq:Hscale}
H_a := \sum_{j=1} ^N \left( \left(-\im \nabla_{\bx_j} - \frac{B}{2} \bx_j^{\perp}\right) ^2 - B \right)+ \sum_{1\leq j < k \neq N} v_a(\bx_j - \bx_k) 
\end{equation}
where the potential 
\begin{equation}\label{eq:vscale}
v_a (\bx) = a^{-2} v(a^{-1} \bx) 
\end{equation}
is scaled to be strong and short-range in the limit $a\to 0$. The convention is that the integral of $v_a$ is fixed, so that the potential converges to a Dirac delta function. The following is proved in~\cite{SeiYng-20} (to which we refer for more comments):

\begin{theorem}[\textbf{Derivation of Haldane pseudo-potentials}]\label{thm:Haldane}\mbox{}\\
 Let $\ell$ be an odd number and the scattering coefficient  
 $$ b_\ell := \frac{1}{4\pi \ell} \min \left\lbrace \int_{\R^2} |\bx|^{2\ell}\left( |\nabla f (\bx)|^2 + \frac{1}{2} v(\bx) |f(\bx)|^2 \right)d\bx,\: f(\bx) \underset{|\bx| \to \infty}{\to} 1 \right\rbrace.$$
Set
$$ c_\ell := 8\pi \ell \left(\pi 2^{\ell +1} \ell!\right)^{-1} b_\ell.$$
When $a\to 0$
$$a^{-2\ell}H_a \to  c_\ell P_{\ell-2,N} H (\ell,N) P_{\ell-2,N}$$ 
in strong resolvent sense\footnote{$H_n \to H$ in this sense if $\left(\mu + H_n\right)^{-1} \psi \to \left(\mu + H\right)^{-1} \psi$ for any state vector $\psi$ and any $\mu >0$.} on $L_{\mathrm{asym}}^2 (\R^{2N})$. Here $H(\ell,N)$ is as in~\eqref{eq:pseupot} and $P_{\ell-2,N}$ projects on the kernel~\eqref{eq:PsiF} of $H(\ell-2,N)$, with the convention $P_{\ell-2,N}= \1$ for $\ell = 1$. 
\end{theorem}

This says that the $\ell$-th Haldane pseudo-potential is obtained at energies of order $a^{2\ell}$ in the limit of a potential of short range $a$. There is a multi-scale aspect: to reach such energies, one must first cancel the $\ell - 2$ first Haldane pseudo-potentials, whence the projection on their kernel. Note that for $\ell = 1$ there is no such lower Haldane pseudo-potential. Hence $a^{-2}H_a$ converges to $H (1,N)$ acting on the whole $\LLLf$ (that one can identify with~\eqref{eq:PsiF} for $\ell = 1$), with ground state space generated from $\PsiLau^{(3)}$ as in~\eqref{eq:PsiF}. Noteworthily, one can identify $H (1,N)$ using derivatives of the delta- interaction potential, 
$$
H(1,N)= \pi  \sum_{i<j} \Delta \delta(z_i-z_j)
$$
where $\delta(z_i-z_j)$ acts as evaluation on the diagonal $z_i= z_j$ (which is a perfectly well-defined operation on the very regular LLL wave-functions).

Corollaries of the above result include that if a suitably small trapping term is added to $a^{2\ell} H_a$, its ground state converges to $\PsiLau^{(\ell)},$ see~\cite{LewSei-09,SeiYng-20} for more details.

\subsection{The spectral gap conjecture}

Hence, in a well-defined albeit idealized limit, the Laughlin state is a true  ground state. However, the dependence on the particle number of the gap above the zero ground state energy is not known. Thus one cannot take the thermodynamic limit at the same time as the short-range limit, while precisely controling the approximation.

The solution to this problem is an important conjecture. It says that the gap above the eigenvalue $0$ does not close in the thermodynamic limit $N\to \infty$. To formulate this, observe first that 
$H(\ell-2,N)$ commutes with the total angular momentum operator 
\begin{equation}\label{eq:tot mom}
\cL_N := \sum_{j=1} ^N z_j \partial {z_j} - \overline{z}_j \partial_{\overline{z}_j}, 
\end{equation}
and consider a joint diagonalization of the two operators on $\LLLf$. The angular momentum of the Laughlin state~\eqref{eq:PsiLau} is
$$ \cL_ N \PsiLaun = \frac{\ell}{2} N (N-1) \PsiLaun.$$

\begin{conjecture}[\textbf{Spectral gap conjecture}]\label{conj:spec}\mbox{}\\
Consider the spectral gap of $H_{\ell-2,N}$ on the sector of angular momenta below that of the Laughlin state
\begin{equation}\label{eq:spectral gap}
 \sigma (N,\ell) = \inf \left\{ \mathrm{spec} \left(H(\ell-2,N) _{|\cL_N \leq \frac{\ell}{2} N (N-1)} \right) \setminus \{ 0 \}\right\}.
\end{equation}
There exists a constant $c_{\ell}>0$, independent of $N$, such that 
$$ \sigma (N,\ell) \geq c_\ell >0.$$
\end{conjecture}

The above is widely believed to be true (and has been advertized by experts) on the grounds that: 

\medskip

\noindent\textbf{1}. It is supported by numerical simulations (numerical diagonalizations of the Hamiltonian for small particle numbers, say up to $\sim 20$, see for example~\cite{Jain-07,Viefers-08} and references therein).

\medskip

\noindent\textbf{2}. Where it to be false, it would be extremely hard to make sense of the experimental data of the FQHE, obtained for thermodynamically large systems.

\medskip 

It should not actually be necessary to restrict the Hamiltonian to angular momenta below $\ell N(N-1)/2$ to obtain a lower bound to the spectral gap. It is likely that restricting to angular momenta below a larger value (but still of order $N^2$ when $N\to \infty$) would suffice. It is conceivable~\cite{YanHalRez-01} that the conjecture holds only for moderate values of $\ell$. 

There are other versions of the conjecture: for particles living on a sphere or a cylinder instead of in the plane, see~\cite[Sections~3.10 and 3.11]{Jain-07} and references therein. The (appropriately modified) conjecture is known to hold~\cite{NacWarYou-20a,NacWarYou-20b,WarYou-21,WarYou-21b} in one such cases: particles confined to a thin torus, a limit in which the problem starts being reminiscent of a 1D quantum spin chain.

\subsection{Stability of the Laughlin phase}\label{sec:stab}

We now take for granted that, in the FQHE regime around $\nu = \ell^{-1}$, leading energy considerations force us to restrict to trial states of the form~\eqref{eq:PsiF}, for they exhaust all the zero-energy eigenstates of the leading Haldane pseudo-potentials. As we discussed above, that is not the end of the story: we now need to justify the further restriction to the simpler states of the form~\eqref{eq:Psif}.

We thus consider a Hamilton function 
\begin{equation}\label{eq:start Hamil}
 \R^{2N} \ni \left( \bx_1,\ldots,\bx_N \right) \mapsto \sum_{j=1}^N V(\bx_j) + \lambda \sum_{1\leq i < j \leq N} W (\bx_i - \bx_j) 
\end{equation}
where $V,W:\R^2 \mapsto \R$ are respectively a one-body and a two-body potential and $\lambda\in \R$ is a coupling constant. We shall discuss the following problem 
\begin{equation}  \label{eq:qm_energy}
E (N,\lambda)=\inf\Big\{\mathcal{E}_{N,\lambda}[\Psi_F]\;|\;\Psi_F \mbox{ of the form~\eqref{eq:PsiF}},\,\int_{\mathbb{R}^{2N}}|\Psi_F|^2=1\Big\}
\end{equation}
where 
\begin{equation} \label{eq:many_body_energy}
\mathcal{E}_{N,\lambda}[\Psi_F]=\Big\langle\Psi_F\Big|\sum_{j=1}^N V(x_j)+\lambda \sum_{i<j}W(x_i-x_j)\Big|\Psi_F\Big\rangle_{L^2}.
\end{equation}
In the spirit of degenerate perturbation theory (again), what the above means is that we minimize the remaining potential energy within the (approximate) ground eigenspace of the main energy scales. The two parts of the Hamilton function represent e.g. energies due to impurities in the sample and/or external fields, and the residual, long-range part of the interaction energy that is not killed by restricting to trial states of the form $\Psi_F$. 

As discussed above, it is important in Laughlin's theory that one can further restrict variational states to the simpler form~\eqref{eq:Psif}. This can be interpreted as the absence of superfluous correlations, and/or the emergence of quasi-holes generated by the action of external fields on the Laughlin ``vacuum''. To formulate this mathematically, define a restricted infimum by setting 
\begin{equation} \label{eq:qh_energy}
e (N,\lambda)=\inf\Big\{\mathcal{E}_{N,\lambda}[\Psi_f]\;|\;\Psi_f \text{ of the form \eqref{eq:Psif}},\,\int_{\mathbb{R}^{2N}}|\Psi_f|^2=1\Big\}.
\end{equation}
Obviously $E(N,\lambda) \leq e(N,\lambda)$. What we would like to prove is that there is equality in the thermodynamic limit:
\begin{equation} \label{eq:anticipation}
\boxed{E (N,\lambda)\simeq e (N,\lambda)\quad\text{as}\;N\to\infty \mbox{ with } \lambda \mbox{ fixed}.}
\end{equation}
We now set the (presumably non-optimal, but illustrative) assumptions under which the above has been proved in~\cite{LieRouYng-16,LieRouYng-17,RouYng-17,OlgRou-19}. Since functions from our variational space~\eqref{eq:PsiF} naturally live over thermodynamically large length scales $\sim \sqrt{N}$ it is natural to scale the potentials $V$ and $W$ accordingly. We thus set, for fixed functions $v,w$,
\begin{equation}\label{eq:rescaled_V}
V (x) = v\left(N^{-1/2} x\right)
\end{equation}
and (the $N^{-1}$ pre-factor ensures that the potential and interaction energies stay of the same order when $N\to \infty$)
\begin{equation} \label{eq:rescaled_w}
W (x) = N ^{-1} w \left(N^{-1/2} x\right).
\end{equation}

\begin{theorem}[\textbf{Energy of the Laughlin phase}]\label{thm:ener}\mbox{}\\
Assume that $v$ and $w$ are smooth fixed functions. Assume that $v$ goes to $+\infty$ polynomially at infinity, and that it has finitely many non-degenerate critical points. There exists $\lambda_0 >0$ such that
$$ \frac{E(N,\lambda)}{e(N,\lambda)} \underset{N\to \infty}{\to} 1$$
with $B > 0$ fixed, $\ell>0$ a fixed integer and $|\lambda| \leq \lambda_0$. 
\end{theorem}

In scaling the external potential as in~\eqref{eq:rescaled_V} we make it live on the natural, thermodynamically large, length-scale of the Laughlin function. This is very reasonable for the trapping part of the potential, but much less so for the part modeling disorder, which typically lives on a much shorter length scale. We can in fact allow shorter length scales, but improving this to realistic values\footnote{The optimal assumption should be that the length scale be much larger than the magnetic length $B^{-1/2}$.} remains an open problem. We prefer to use a single length scale, in order not to obscure the statement. 

In Theorem~\ref{thm:ener} we assume the interaction to be smooth. This is because it is supposed to represent the long-range part only, the singular short-range part being taken care of by restricting to~\eqref{eq:PsiF}. Scaling $W$ as in~\eqref{eq:rescaled_w} has the merit of making the two terms in~\eqref{eq:many_body_energy} of the same order of magnitude, as in a mean-field limit. This also simplifies statements a lot, but for interactions scaling like 3D Coulomb, this is actually the correct thing to do, see~\cite[Section~2.2]{OlgRou-19}.

Concerning the smallness assumption on $\lambda$ in~Theorem~\ref{thm:ener}, it corresponds to the fact that the filling factor should stay close to $\ell^{-1}$ for the theorem to be true. Too large a deviation makes the system jump to a different FQHE plateau, e.g. a Laughlin state with higher exponent. Increasing the (repulsive) interaction strength has the net effect of spreading the system further, and hence lowering the filling factor (see again~\cite[Section~2.2]{OlgRou-19} for more details). An upper bound, probably model-dependent and hard to estimate, on $|\lambda|$ is thus necessary for the statement to hold. 

\subsection{Incompressibility estimates}

We will not go into the details of the proof of Theorem~\ref{thm:ener} (which also has corrolaries regarding the optimal densities). We only mention that it mostly relies on a remarkable rigidity property shared by all states of the form~\eqref{eq:PsiF}. We coined this an ``incompressibility estimate'' in~\cite{RouYng-14,RouYng-15} where partial results were obtained before the full result below was proved in~\cite{LieRouYng-17} and improved in~\cite{OlgRou-19}. This notion of incompressibility should not be confused with that discussed in Section~\ref{sec:gap}, that it complements (and of which it is logically independent).

Let 
\begin{equation}\label{eq:density}
\varrho_{\Psi_F} (\bx) := N \int_{\R^{2(N-1)}} \left| \Psi_F (\bx,\bx_2,\ldots,\bx_N) \right|^2 d\bx_2 \ldots d\bx_N
\end{equation}
be the one-particle density associated to $\Psi_F$ of the form~\eqref{eq:PsiF}. We have the universal bound

\begin{theorem}[\textbf{Incompressibility estimate}]\label{thm:incomp}\mbox{}\\ 
For any $\alpha>(\sqrt{5}-1)/4$, any disk  $D$ of radius $N^{\alpha}$ and any (sequence of) states $\Psi_F$ of the form~\eqref{eq:PsiF} we have 
\begin{equation}\label{bound}
\int_{D} \varrho_{\Psi_F} \leq \frac{B}{2\pi\ell} |D| (1+o(1))
\end{equation}
where $|D|$ is the area of the disk and $o(1)$ tends to zero as $N\to \infty$. 
\end{theorem}

What this says is that, in the sense of averages on sufficiently large length scales (which are allowed to be much smaller than the thermodynamic length scale $\sim \sqrt{N}$) and independently of $F$  
$$\varrho_{\Psi_F} \lessapprox \frac{B}{2\pi\ell}.$$
I.e. the local filling factor can, in states of the form~\eqref{eq:PsiF}, nowhere be larger than the global filling factor of the Laughlin state.

We conjecture that the result only requires $N^\alpha \gg B^{-1/2}$ (i.e. that the bound holds in the sense of averages on any length scale larger than the magnetic length), but this remains an open problem. Its' solution would go a long way towards removing assumptions on length scales made in Theorem~\ref{thm:ener}.

The proof of the above is based on Laughlin's plasma analogy, which maps the $N$-particle density of $\Psi_F$ to Gibbs states of fictitious classical 2D electrostatic systems (somewhat contrived for non trivial $F$). Screening considerations valid in a large generality for these systems yield the result. The form of $F$ can a priori be quite wild, but, being analytic, it leads to a repulsive electrostatic force in the plasma analogy. This can only lower the density compared to the case with no $F$. Most of the difficulties lie in vindicating this intuitive (but specific to Coulomb interactions) fact. One needs to show it for essentially any form of the corresponding charge distribution, which can be quite bizarre and correlated for general $F$.

\section{Conclusion}

We gave a very rough sketch of Laughlin's theory, the most basic building block of our theoretical understanding of the fractional quantum Hall effect. We argued that two crucial properties, assumed in the general theory, play a key role in making it an efficient description. From a mathematical physics point of view, the first property (``incompressibility'') rests on Haldane pseudo-potentials and an open problem concerning their spectra, the spectral gap conjecture (partial results were however obtained recently~\cite{NacWarYou-20a,NacWarYou-20b,WarYou-21,WarYou-21b}). The second property (``rigidity'') can be given the form of a stability question for Laughlin quasi-holes generated from the vacuum of the theory. This has been solved in some generality recently, although important questions remain on the optimal assumptions allowed in the current state of affairs. We did not mention at all extensions of the above considerations to other quantum Hall fractions than the inverse odd integers, simply because this remains a wide open problem. 

\bigskip

\textbf{Acknowledgments:} Thanks to Tapash Chakraborty for inviting me to write this contribution and to Jakob Yngvason for helpful remarks. I am financially supported by the European Research Council (under the European Union's Horizon 2020 Research and Innovation Programme, Grant agreement CORFRONMAT No 758620). 

\newpage
% % 
% \bibliographystyle{acm}
% \bibliography{/home/rougerie/Travail/Documentation/Bibtex/biblio-NR_Aug22.bib}
% 

\end{document}